# PID Tuning via Desired Step Response Curve Fitting

Senol Gulgonul
*Electrical and Electronics Engineering, Ostim Technical University, Ankara Turkey*
*senol.gulgonul@ostimteknik.edu.tr*

**Abstract:** This paper presents a PID tuning method based on step response curve fitting (PID-SRCF) that utilizes L2-norm minimization for precise reference tracking and explicit transient response shaping. The algorithm optimizes controller parameters by minimizing the root-mean-square error between desired and actual step responses. The proposed approach determines optimal PID parameters by matching any closed-loop response to a desired system step response. Practically a first-order plus time delay model or a second-order system with defined settling time and overshoot requirements are preferred. The method has open-source implementation using constrained nonlinear optimization in MATLAB. Comparative evaluations demonstrate that PID-SRCF can replace known analytical methods like Ziegler Nichols, Lambda Tuning, Pole Placement, Dominant Pole and MATLAB proprietary PID tuning applications.

## 1. Introduction

PID controllers have dominated control engineering since their early development in the 1870s-1920s, with their modern formulation emerging during the 1930s-1940s process control revolution [1]. Their industrial prevalence remains unparalleled: a 1989 Japanese study revealed over 90% of control loops utilized PID control, while a 2002 U.S. survey identified PID controllers in 97% of regulatory loops across petrochemical, power, and manufacturing industries - totaling millions of installations [2,3]. The technology's scalability is evidenced by its deployment in precision applications like optical drives (140 million units annually for CD/DVD tracking systems) [4] and ubiquitous motor control implementations. This enduring dominance, spanning basic process regulation to high-volume consumer electronics, solidifies PID control as the de facto standard for industrial automation.

Tuning of PID controller problem can be defined as the calculation of PID coefficients for desired performance requirements for the given process system. Fundamental performance requirements in time domain are based on step response of the system such as settling time, percentage of overshoot. In frequency domain gain margin and phase margin are two main requirements. Besides, robustness, load disturbance attenuation and measurement noise response have to be considered [5].

The foundational work of Ziegler and Nichols (1942) established the first systematic PID tuning methodology by experimentally determining that optimal control could be achieved by targeting a 25% amplitude ratio in the closed-loop step response. Their closed-loop method derived controller parameters by first identifying the ultimate gain Ku and oscillation period Pu at the stability limit, then setting the proportional gain to 0.5Ku and integral time to Pu/1.2 to achieve the desired damping. Their complementary open-loop method used the process reaction curve's lag time L and reaction rate R to calculate equivalent settings, where the characteristic S-shaped response and time-domain parameters (L, R) implicitly reflected first-order-plus-time-delay (FOTD) dynamics, though no explicit plant model was specified. Both approaches target 25% amplitude decay while accommodating different experimental conditions [6].

The Cohen-Coon method, based on the FOTD model optimizes PID tuning by positioning dominant poles to achieve a quarter-amplitude decay ratio (25% amplitude ratio in the closed-loop step response). For PID control, two complex poles (with damping ratio $\zeta \approx 0.215$) and one real pole are placed at equal distances from the origin, maximizing integral gain to minimize integral error (IE) for step disturbances while maintaining the target 25% amplitude decay [7].

Åström and Hägglund's (1984) relay auto-tuning method automatically identifies process dynamics by inducing limit cycle oscillations through relay feedback, measuring ultimate gain (Ku) and period (Pu). The method preserves Ziegler-Nichols' 25% amplitude decay ratio criterion while eliminating manual testing, providing reliable PID tuning through sustained oscillations that match classical stability targets. This innovation enabled automated implementation of established design specifications [8].

Lambda tuning, first suggested by Dahlin (1968), is a pole-placement method designed for first-order plus time-delay (FOTD) plants characterized by a time constant T and delay L [9]. The method aims to achieve a closed-loop response that mimics a FOTD system, where the controller cancels the process pole by setting the integral time Ti=T for a PI controller. This cancellation, however, can lead to poor load disturbance rejection in lag-dominated systems [10].

To address this limitation, an alternative approach avoids pole-zero cancellation by treating the closed-loop system as second-order. Here, Ti ≠ T, and the controller parameters are derived to place both poles at the desired closed-loop time constant $\lambda$. This modification improves disturbance rejection while retaining the simplicity of the Lambda method, which requires only one tuning parameter ($\lambda$) [10].

Åström introduced the maximum sensitivity ($M_s$) as a robust tuning parameter for PID controller design. Geometrically, $M_s$ corresponds to the inverse of the minimal distance from the Nyquist curve of the loop transfer function to the critical point (-1,0). This provides a direct measure of robustness, as it quantifies how much the process dynamics can vary before closed-loop instability occurs. Typical $M_s$ values range from 1.2 to 2.0, where lower values yield more conservative control with greater stability margins, while higher values provide faster response at the cost of reduced robustness. While the $M_s$-method provides direct control over robustness through sensitivity function shaping, it lacks an explicit analytical relationship to time-domain performance metrics like overshoot and settling time, requiring iterative refinement or complementary techniques when strict transient response specifications must be met [10].

Åström demonstrated that unconstrained IAE optimization for constant setpoints leads to impractical controller designs with excessive gain amplification and poor robustness. This necessitated the introduction of $M_s$ constraints to achieve stable implementations. While these results confirmed that perfect constant setpoint tracking is theoretically unachievable in real systems, they inspired our development of an alternative approach: applying L2-norm minimization to time-varying reference trajectories. This method allows the target system response to be specified either manually or as the step response of a user-defined transfer function (e.g., FOPTD, second-order, or other dynamics), providing more flexible and realistic performance objectives for controller tuning [11].

Model Predictive Control (MPC) is an advanced control strategy that employs a receding horizon optimization framework to compute control actions by solving a constrained optimization problem at each sampling instant [12,13]. Utilizing an explicit dynamic model of the process, MPC minimizes a cost function—typically penalizing tracking error and control effort—over a finite prediction horizon while respecting system constraints such as actuator limits and state bounds. Its ability to handle multivariable systems, nonlinear dynamics, and hard constraints has made it indispensable in industries ranging from chemical processing to automotive control [14]. Unlike conventional PID control, MPC explicitly accounts for future system behavior and adapts to time-varying references and disturbances through online optimization. In contrast to the proposed PID-SRCF algorithm, which optimizes PID parameters offline for predefined target trajectories, MPC performs real-time optimization, offering greater flexibility for dynamic environments but at higher computational cost. This distinction positions the PID-SRCF method as a computationally efficient alternative for systems where offline tuning suffices and explicit trajectory tracking is prioritized.

MATLAB implements a proprietary frequency-domain PID tuning algorithm (patented) that is accessible either through the command-line *pidtune* function or its graphical *Control System Tuner* and *PID Tuner app* [15]. *Control System Tuner* tunes the PID parameters to best meet the tuning goals (design requirements). Tuning goals include reference tracking, disturbance rejection, loop shapes, closed-loop damping, and stability margins. In this study, MATLAB's *Control System Tuner* function serves as a representative industrial-grade PID tuning benchmark for comparison with the developed PID-SRCF method for high order systems.

## 2. Methodology

### 2.1 Overview

The Step Response Curve Fitting PID (PID-SRCF) tuning algorithm presents a generalized methodology through direct minimization of the L2-norm error between the closed-loop response and a desired system step response. This numerical optimization approach, implemented as an open-source MATLAB script, offers control engineers an alternative to conventional tuning techniques including Ziegler-Nichols, Lambda tuning, pole placement, and dominant pole methods or proprietary tuning tools like MATLAB *Control System Tuner*.

The PID-SRCF algorithm initiates by specifying a desired closed-loop response, which can be defined either as a time series or, more practically, through a transfer function representation. For systems requiring non-oscillatory behavior, a first-order plus time delay (FOTD) transfer function serves as an effective target, replacing conventional lambda tuning methods. This first-order target response formulation remains applicable to processes of arbitrary order when the design objective is to achieve stable, first-order-like dynamics in the closed-loop system. The method provides particular advantages when smooth, non-oscillatory transient responses are desired, while maintaining the flexibility to accommodate higher-order process dynamics through its numerical optimization framework.

The algorithm supports second-order desired responses defined through two key performance parameters: settling time ($T_s$) and percent overshoot (PO). For non-oscillatory responses, a critically damped target is automatically generated, while underdamped systems incorporate calculated damping characteristics based on the specified overshoot. PID-SRCF approach subsumes conventional tuning objectives from both industrial methods like Ziegler-Nichols and academic approaches such as pole placement, while providing unified control over transient response characteristics through intuitive performance parameters. The second-order formulation offers particular advantages when either precise overshoot control or balanced response times are required, maintaining flexibility across different control applications.

The PID parameter optimization is formulated as a constrained minimization problem that adjusts the controller coefficients ($K_p$, $K_i$, $K_d$) to minimize the L2-norm error between the achieved and desired step responses. The numerical solution employs Sequential Quadratic Programming (SQP) to handle the nonlinear optimization using MATLAB *fmincon* to find minimum of constrained nonlinear multivariable function. While the primary optimization objective focuses on time-domain tracking performance, the algorithm includes post-optimization evaluation of the maximum sensitivity ($M_s$) to assess robustness margins. The target response specification inherently promotes robustness through appropriate dynamic characteristics, though the framework readily accommodates explicit $M_s$ constraints through straightforward modification of the optimization constraints. The current design preserves the method's simplicity while providing flexibility for engineers to enforce additional robustness requirements when necessary.

Comparative analysis with MATLAB's *Control System Tuner* demonstrates the PID-SRCF algorithm's ability to achieve almost the same PID coefficients for non-overshooting requirements. The offline optimization approach maintains computational efficiency while accommodating desired transfer functions of arbitrary order, though in practice, first-order and second-order reference models prove most practical for control applications. Complete implementation details and theoretical foundations are presented in subsequent sections.

### 2.2 Desired System Design

The desired system can be defined by the user as a transfer function. For second order underdamped systems with specified percent overshoot (PO), the damping ratio (ζ) is calculated as:

$$\zeta \approx -\frac{\ln\left(\frac{PO}{100}\right)}{\sqrt{\pi^2 + ln^2\left(\frac{PO}{100}\right)}} \tag{1}$$

where PO represents the permissible percentage overshoot. The corresponding natural frequency ($w_n$) for a 2% settling time ($T_s$) criterion is determined by:

$$w_n \approx \frac{4}{\zeta T_s} \tag{2}$$

yielding the standard second order transfer function:

$$G_{desired}(s) = \frac{{w_n}^2}{s^2 + 2\zeta w_n s + {w_n}^2} \tag{3}$$

For critically damped systems (PO = 0%), the design simplifies with ζ = 1 and modified natural frequency:

$$w_{nc} \approx \frac{6}{\zeta T_s} \tag{4}$$

resulting in the non-oscillatory and non-overshooting desired system

$$G_{desired}(s) = \frac{{w_{nc}}^2}{s^2 + 2w_{nc}s + {w_{nc}}^2} \tag{5}$$

A general form of the desired first order systems with time delay (FOTD) can be defined as

$$G_{desired}(s) = \frac{1}{1 + sT_{cl}} e^{-sL} \tag{6}$$

where $T_{cl}$ time constant of desired closed loop system and L is delay. Delay L has to be same as the process transfer function. FOTD system will generate again a non-overshooting step response and settling time for 2% criteria will be

$$T_s \approx 4T_{cl} \tag{7}$$

### 2.3 Optimization Problem

The controller design problem is formulated as a constrained optimization task minimizing the $L_2$-norm error between the closed-loop step response $y_{PID}(t)$ and the desired reference trajectory $y_{desired}(t)$. The objective function is defined as:

$$\min_{K_p, K_i, K_d} \sqrt{\int_0^{T_{sim}} \left|\left(y_{desired}(t) - y_{PID}(t)\right)\right|^2 dt} \tag{8}$$

The optimization is implemented numerically using MATLAB's *fmincon* solver with the Sequential Quadratic Programming (SQP) algorithm. Initial conditions start from zero ($K_p$=$K_i$ =$K_d$= 0) and constraints of non-negative PID gains. Upper bounds are left to user to define according to type of PID controller such as P, PI, PD or PID.

$$K_p, K_i, K_d \geq 0 \tag{9}$$

Closed-loop stability is verified ex post by inspecting pole locations of the tuned system:

$$Re(poles) < 0 \tag{10}$$

Performance metrics include IAE, settling time, overshoot, and maximum sensitivity $M_s$, computed via the sensitivity function

$$Ms = max \mid \frac{1}{1 + L(j\omega)} \mid = max \mid S(j\omega) \mid \tag{11}$$

where $L(j\omega)$ is the open-loop transfer function.

## 3. MATLAB Implementation

The PID controller optimization was implemented in MATLAB environment using Control System Toolbox and Optimization Toolbox. The implementation architecture comprises four main components: desired response generation, optimization framework, stability verification, and performance evaluation.

The PID-SRCF algorithm demonstrates efficient computational performance, with execution times consistently below one second when implemented on a standard notebook computer (Intel Core i7-2.8GHz processor, 16GB RAM).

As an example, the desired second-order system is generated programmatically based on design specifications. For systems requiring zero overshoot (critically damped), the transfer function is constructed with damping ratio $\zeta = 1$ and natural frequency $wn \approx 6/T_s$:

```
if PO==0  % critically damped
   zeta=1;
   wn = 6/(zeta*Ts);
   desired_sys = wn^2/(s^2 + 2*zeta*wn*s + wn^2);
else   % underdamped
   zeta = -log(PO/100)/sqrt(pi^2 + log(PO/100)^2);
   wn = 4/(zeta*Ts);                         % Natural frequency
   desired_sys = wn^2/(s^2 + 2*zeta*wn*s + wn^2);
end
```

The optimization core utilizes MATLAB's nonlinear constrained minimization function *fmincon* with Sequential Quadratic Programming algorithm. The objective function computes the $L_2$-norm error between the desired and achieved step responses, minimizing the root-mean-square deviation for optimal reference tracking.

```
error_func = @(x) norm(desired_response - ...
                    step(feedback((x(1) + x(2)/s + x(3)*s)*G, 1), t), 2);
```

The *fimincon* solver was configured with a maximum of 3000 function evaluations to ensure computational feasibility while maintaining solution accuracy. Convergence progress was monitored through iterative display output. Parameter constraints were enforced by specifying non-negative lower bounds for all PID gains (Kp, Ki, Kd ≥ 0) to ensure physically meaningful controller implementations. The optimization was initialized from zero gain values to avoid bias in the solution search.

```
options = optimoptions('fmincon', ...
   'Algorithm', 'sqp', ...
   'MaxFunctionEvaluations', 3000, ...
   'Display', 'iter');
solution = fmincon(error_func, [0; 0; 0], [], [], [], [], ...
            [0; 0; 0], [Inf; Inf; Inf], [], options);
```

Stability verification is performed through analytical examination of closed-loop pole locations.

```
poles = pole(T);
is_stable = all(real(poles) < 0)
```

The implementation includes comprehensive comparison with results presented in tabular format showing controller parameters ($K_p$, $K_i$, $K_d$), settling time ($T_s$), percent overshoot (PO%), IAE values, and maximum sensitivity ($M_s$). Visual comparison is generated through superimposed step response plots.

```
C = Kp + Ki/s + Kd*s;        % PID controller
L = G * C;                   % Open-loop transfer function
S = 1 / (1 + L);             % Sensitivity function
[mag, ~, ~] = bode(S);       % Magnitude across frequencies
Ms_pid_srcf = max(mag);      % Peak sensitivity for pid_srcf
```

## 4. Results and Discussion

PID-SRCF method compared to Ziegler Nichols, Lambda, Pole Placement, Dominant Pole and MATLAB *Control System Tuner* PID tuning methods with examples.

### 4.1. Ziegler-Nichols Case

We will go through the example in Astrom's Advanced PID Control book with process transfer function [5]:

$$G(s) = \frac{1}{(s+1)^3} \tag{12}$$

Ziegler-Nichols criteria for one quarter decay ratio $d = 0.25$ gives damping ratio $\zeta = 0.215$ for a second order system.

$$d = e^{\frac{-2\pi\zeta}{\sqrt{1-\zeta^2}}} \tag{13}$$

The process has ultimate gain $K_u = 8$ with ultimate period $T_u = 3.63$ seconds yields natural frequency

$$w_n = \frac{2\pi}{T_u} = 1.73 \qquad rad/s \tag{14}$$

Finally, the desired second order system which satisfies Ziegler-Nichol's criteria

$$G_{desired}(s) = \frac{{w_n}^2}{s^2 + 2\zeta w_n s + {w_n}^2} \tag{15}$$

PID-SRCF with this desired transfer response calculates optimum PID coefficients close to Ziegler Nicols formula for reaction curve method. PID-SRCF also satisfies Ziegler Nichols quarter amplitude decaying criteria. Sensitivity is greater than 2 for both methods which is as expected due to one quarter decay ratio requirement.

**Table 1.** Comparison of PID-SRCF and Ziegler Nichols PID Tuning

| Method | Kp | Ki | Kd | Ts | PO (%) | IAE | Ms |
|---|---|---|---|---|---|---|---|
| pid_srcf | 6.7358 | 3.9912 | 3.0012 | 9.8482 | 49.9890 | 1.8102 | 2.5656 |
| Ziegler Nichols | 5.500 | 3.4200 | 2.2000 | 12.9732 | 51.4006 | 2.1283 | 2.7529 |

Step response curves shows that PID-SRCF method perfectly fits to the desired second order step response with a better settling time. There is time shift compare to Ziegler Nichols PID tuned step response for $w_n = 1.73$ and interestingly for $w_n = 1.54$ both methods fit to desired step response curve. We can conclude that Ziegler Nichols method indirectly desiring a second order closed loop transfer function step response and can be solved by PID-SRCF method.

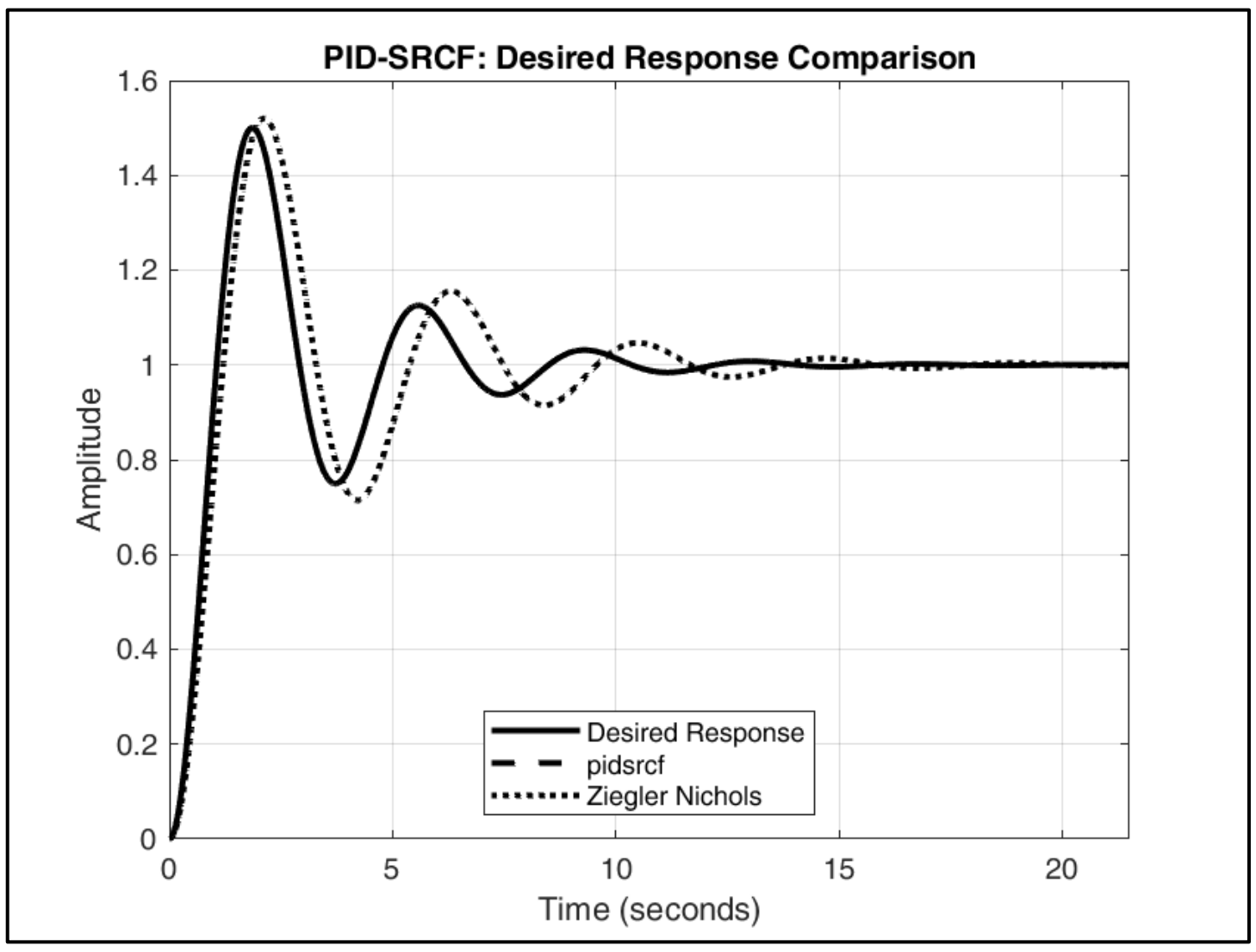

**Figure 1.** Comparison of PID-SRCF and Ziegler Nichol's PID Tuning

### 4.2. Lambda Tuning Case

Lambda tuning example is a FOTD process transfer function having coefficients $K_p = 1$, T=1 and L=1.

$$P(s) = \frac{K}{1 + sT} e^{-sL} \tag{16}$$

In Lambda tuning closed loop system is also expected to be a FOTD transfer function. Closed loop time constant is expected in a range for proper tuning as $T < T_{cl} < 3T$. We selected the midpoint as $T_{cl} = 2T$.

$$G_{cl}(s) = \frac{1}{1 + sT_{cl}} e^{-sL} \tag{17}$$

For PI tuning in Lambda method and pole-zero cancellation

$$K_p = \frac{T}{K(L + T_{cl})} = 0.33 \tag{18}$$

$$K_i = \frac{K_p}{T} = 0.33 \tag{19}$$

PID-SRCF method optimized PI coefficients are very close to Lambda tuning and response curves perfectly fitting means that PID-SRCF can replace Lambda PI tuning with pole-zero cancellation.

**Table 2.** Comparison of PID-SRCF and Lambda PI Tuning

| Method | Kp | Ki | Kd | Ts | PO (%) | IAE | Ms |
|---|---|---|---|---|---|---|---|
| pid_srcf | 0.3955 | 0.3282 | 0.0000 | 9.4693 | 0.0000 | 3.0447 | 1.3568 |
| Lambda PI | 0.3300 | 0.3300 | 0.0000 | 8.0203 | 0.0000 | 3.0303 | 1.3444 |

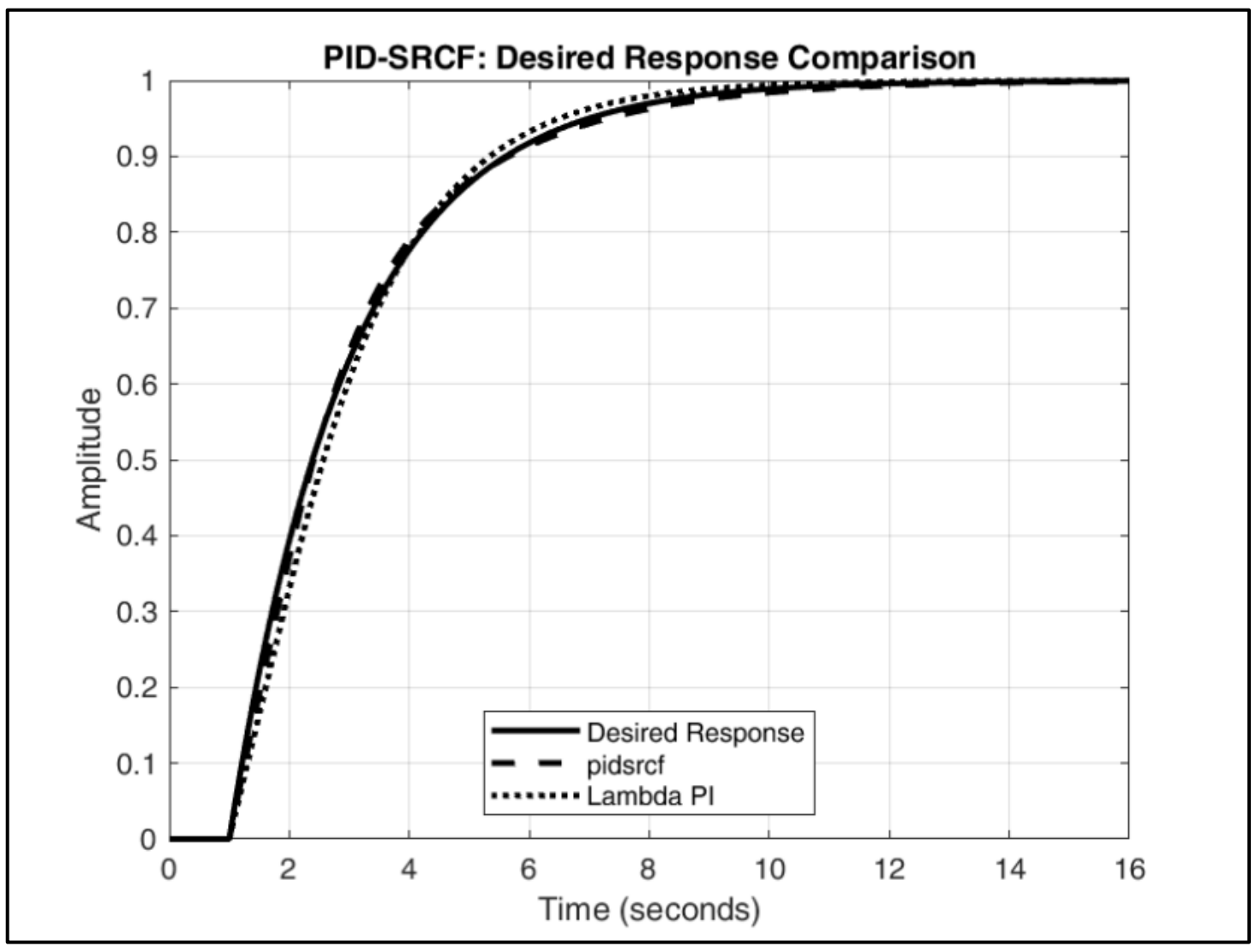

**Figure 2.** Comparison and PID-SRCF to Lambda PI Tuning

### 4.3. Pole Placement or Dominant Pole Case

Consider a first order process system:

$$P(s) = \frac{1}{1 + sT} \tag{20}$$

Closed loop system with PI controller will have two poles and a zero.

$$G_{cl}(s) = \frac{sK_p + K_i}{Ts^2 + (1 + K_p)s + K_i} \tag{21}$$

We can tune PI controller for a second order desired response using PID-SRCF. Let's target a non-overshooting response with Ts=1 seconds. Natural frequency for critically damped second order system is given as

$$w_n \approx \frac{6}{\zeta T_s} = 6 \tag{22}$$

Analytical solution yields PI coefficients for T=1 as

$$K_p = 2\zeta w_n - 1 = 11 \tag{23}$$

$$K_i = {w_n}^2 T = 36 \tag{24}$$

Pole placement for this case does not take transfer function's zero into account, thus resulting behavior is not guaranteed and generating an unexpected overshoot. It is observed that PID-SRCF has lower gains which are preferrable for practical usage.

**Table 3.** Comparison of PID-SRCF and Pole Placement

| Method | Kp | Ki | Kd | Ts | PO (%) | IAE | Ms |
|---|---|---|---|---|---|---|---|
| pid_srcf | 2.5575 | 3.4360 | 0.0000 | 2.1431 | 2.5550 | 0.3454 | 0.9998 |
| Pole Placement PI | 11.0000 | 36.0000 | 0.0000 | 0.8479 | 9.2331 | 0.1115 | 1.0000 |

PID-SRCF overlaps desired critically damped second order step response with a small overshoot of 2.5% which is probably coming approximations of $T_s$, $w_n$ for 2% criteria.

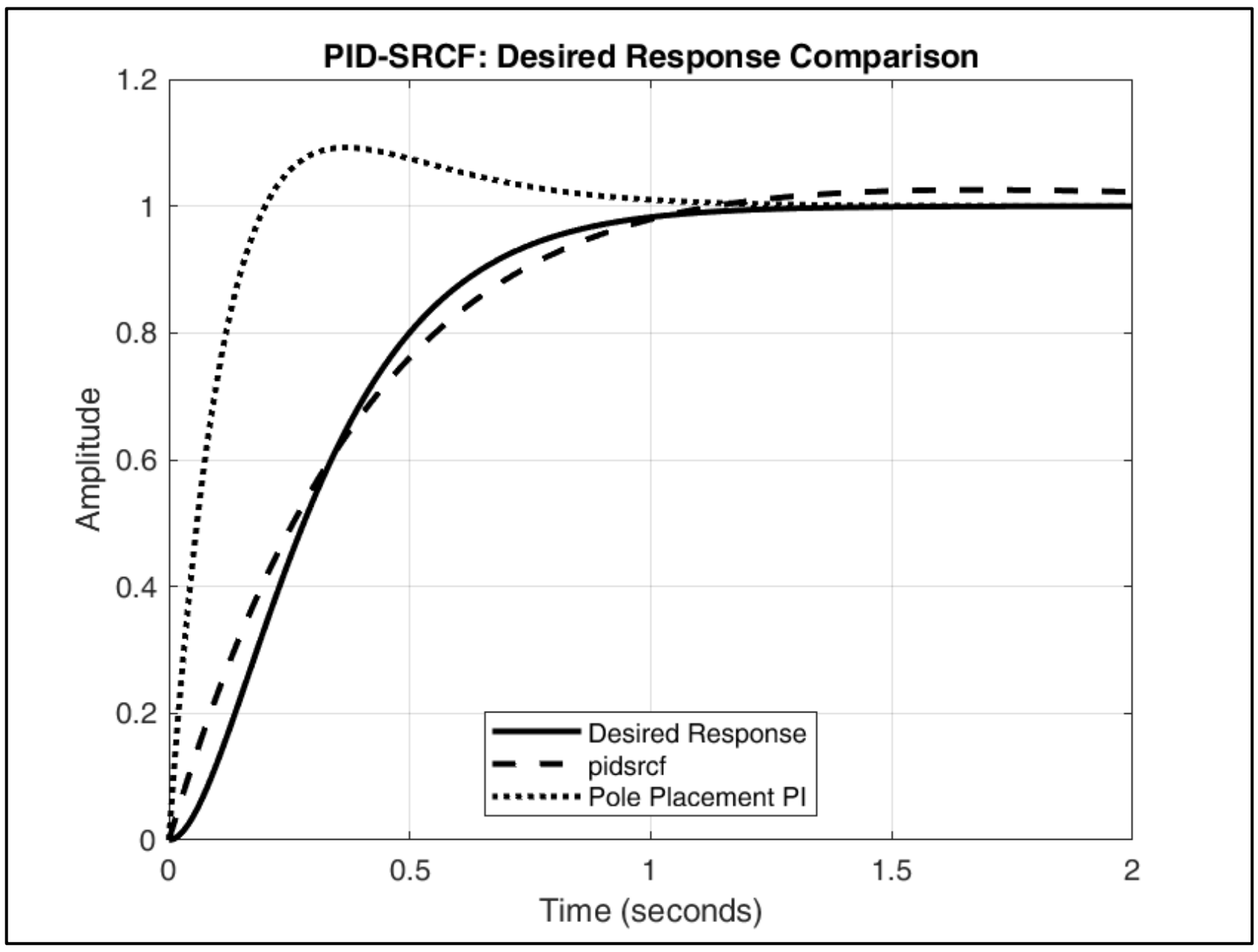

**Figure 3.** Comparison of PID-SRCF and Pole Placement

### 4.4. High Order Transfer Function Case

Final example is PI tuning of a third order process system $P(s)$ for a target first order step response. High order (greater than 2) systems are complex to tune; there is no analytical method and dominant pole approximations does not guarantee the overshoot. Thus, we compared results with MATLAB *Control System Tuner*, with a target zero overshoot first order system with time constant $T = 3$ generated results in Table 4 and PI coefficients are almost the same.

$$G_{desired}(s) = \frac{1}{3s + 1} \tag{25}$$

$$P(s) = \frac{1}{(s + 1)^3} \tag{26}$$

**Table 4.** Comparison of PID-SRCF and MATLAB *Control System Tuner*

| Method | Kp | Ki | Kd | Ts | PO (%) | IAE | Ms |
|---|---|---|---|---|---|---|---|
| pid_srcf | 0.9248 | 0.2829 | 0.0000 | 16.3312 | 0.0000 | 3.5125 | 1.4063 |
| *Control System Tuner App* | 0.9242 | 0.2828 | 0.0000 | 16.3000 | 0.0000 | | |

This was an example to show that PID-SRCF is especially powerful for non-overshooting step response requirements for high order systems as shown in Figure 4.

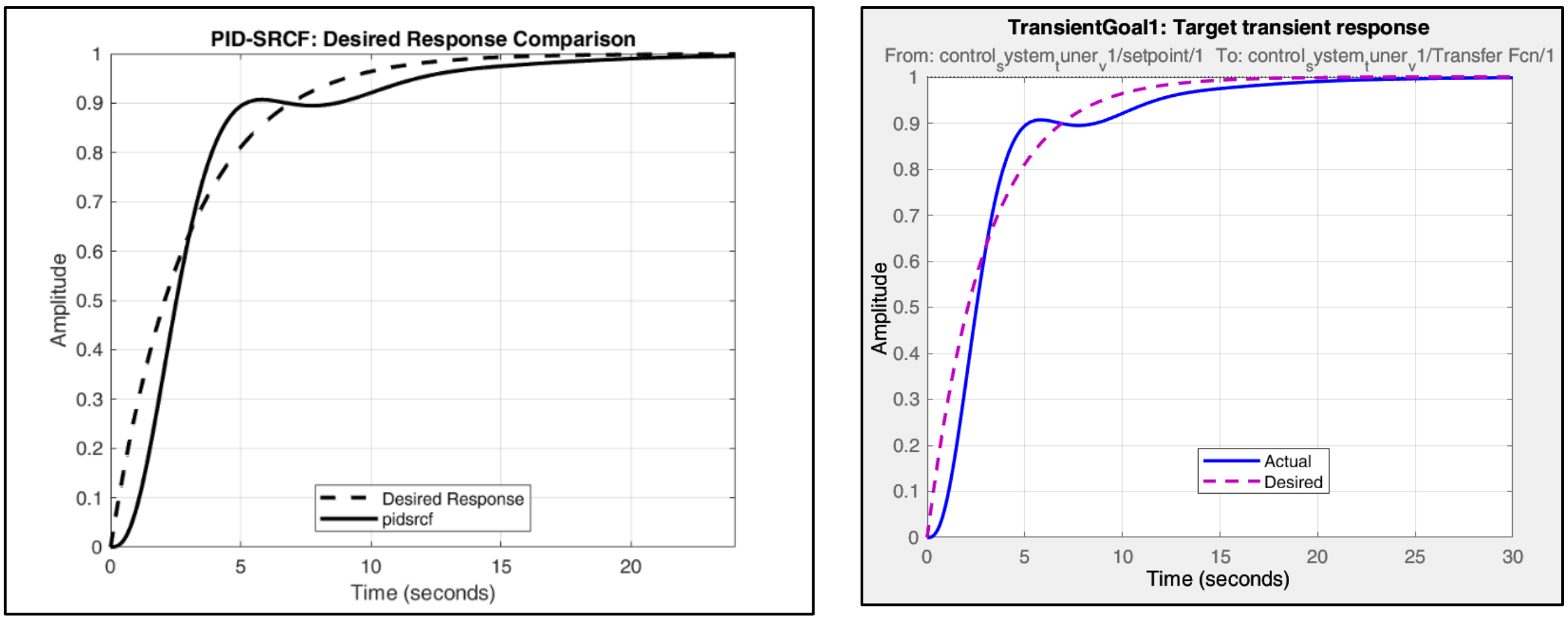

**(a)** **(b)**

**Figure 4.** Comparison of **(a)** PID-SRCF and **(b)** MATLAB *Control System Tuner*

Testing and comparing PID-SRCF with known analytical PID tuning methods; Ziegler Nichols, Lambda tuning, pole placement and dominant pole methods results that PID-SRCF can easily replace these methods. PID-SRCF has also advantageous to MATLAB proprietary PID tuning function *Control System Tuner* due to its open-source flexibility. It was also interesting to see that non-overshooting requirements automatically results maximum sensitivity ($M_s < 2$). This will be presented analytically also in the following chapter.

The four-significant-figure agreement in Table 4 ($K_p = 0.9248$ vs 0.9242, $K_i = 0.2829$ vs 0.2828) is not coincidental. MATLAB's *Control System Tuner* supports a *StepTracking* tuning goal that minimizes the deviation between the closed-loop step response and a user-supplied reference model, which is structurally equivalent to the PID-SRCF formulation of Eq. (8). The patented *systune* solver used by *Control System Tuner* [15] and the open-source *fmincon*/SQP solver used by PID-SRCF converge to essentially the same minimum because the underlying optimization is the same. This option in *Control System Tuner* is not widely known among practitioners, who more commonly use the bandwidth/phase-margin sliders of the *PID Tuner* app or the loop-shaping heuristic of *pidtune*. The contribution of PID-SRCF over the proprietary toolbox is therefore not algorithmic novelty but transparency: a short open-source implementation that any practitioner can inspect, modify, and embed without licensing constraints.

### 4.5. Maximum Sensitivity ($M_s$) for Non-Overshooting Step Response

In this study two non-overshooting desired step response cases are presented: first order and critically damped second order desired closed loop transfer function. For a first order closed loop system maximum sensitivity can be calculated as:

$$T(s) = \frac{K}{sT + 1} \tag{27}$$

$$S(s) = 1 - T(s) = \frac{sT + 1 - K}{sT + 1} \tag{28}$$

$$M_s = max|S(jw)| = \max\left\{\frac{\sqrt{(wT)^2 + (1-K)^2}}{\sqrt{(wT)^2 + 1}}\right\} = 1 \ \ at \ \ w = \infty \tag{29}$$

Similarly for a second order critically damped system we can calculate maximum sensitivity:

$$G_{cl}(s) = \frac{{w_n}^2}{s^2 + 2w_n s + {w_n}^2} \tag{30}$$

$$S(s) = 1 - G_{cl}(s) = \frac{s^2 + 2w_n s}{s^2 + 2w_n s + {w_n}^2} \tag{31}$$

$$M_s = ax|S(jw)| = \max\left\{\frac{w\sqrt{w^2 + 4{w_n}^2}}{\sqrt{({w_n}^2 - w^2)^2 + 4{w_n}^2 w^2}}\right\} = \frac{2}{\sqrt{3}} \approx 1.155 \ \ at\, w = w_n\sqrt{2} \tag{32}$$

Both analytical and empirical results demonstrate that specifying a non-overshooting desired response whether first-order or critically damped second-order inherently guarantees robustness by bounding the maximum sensitivity to $1 < M_s < 2$. This aligns with industrial design thresholds ($M_s < 1.2 - 2.0$) and validates the PID-SRCF method's empirical performance.

The critically damped second-order result $M_s = 2/\sqrt{3}$ of Eq. (32) is the n = 2 case of a more general family of closed-loop transfer functions of the form $T(s) = p^n/(s + p)^n$, for which all classical robustness measures admit universal closed-form values depending only on n and not on the pole location p. The companion paper [16] derives $M_s = 9/7 \approx 1.286$ for n = 3, $M_s \approx 1.381$ for n = 4, together with the closed-form gain margin $GM(n) = 1 + \sec^n(\pi/n)$ and the corresponding phase margins. These constants provide a precise frequency-domain anchor for the desired step response chosen in PID-SRCF: when the target is $T(s) = \omega_n^2/(s + \omega_n)^2$, the user-specified settling time determines $\omega_n$, and the target inherits the universal values $M_s = 2/\sqrt{3}$, $M_t = 1$, PM = 76.35°, and GM = ∞. The realized values in the curve-fit closed loop track these target values up to the $L_2$ fitting residual, with the realized $M_s$ itself serving as a diagnostic of target feasibility for the given plant and controller order.

### 4.6. PID-SRCF as a Unifying Framework

The case studies in Sections 4.1–4.4 reveal a common structure: every classical PID tuning method implicitly chooses a target closed-loop response shape and produces controller gains that realize that shape on the given plant. The methods differ only in which target they pick and how transparently they express that choice. Ziegler-Nichols implicitly targets a quarter-decay second-order response; Lambda tuning explicitly targets a FOTD model; pole placement targets a critically damped second-order; and the companion analytical paper [16] targets the universal-robustness family $T(s) = p^n/(s + p)^n$. PID-SRCF is the meta-method in which the target choice becomes an explicit user input, recovering each classical method as a special case for the corresponding $G_{desired}(s)$. Table 5 summarises the mapping.

**Table 5.** Classical PID tuning methods as special cases of PID-SRCF.

| Method | Implicit target G_desired(s) | User parameters |
|---|---|---|
| Ziegler-Nichols (closed-loop) | $\omega_n^2/(s^2 + 2\zeta\omega_n s + \omega_n^2)$, $\zeta \approx 0.215$, $\omega_n = 2\pi/Tu$ | Ku, Tu from relay test |
| Cohen-Coon | Same family as Ziegler-Nichols; quarter-decay second-order target with different ω_n estimation | K, L, T from step test |
| Lambda tuning (FOTD) | $e^{(-sL)}/(1 + s\,Tcl)$ | T_cl (closed-loop time constant) |
| IMC tuning | $e^{(-sL)}/(1 + s\lambda)^n$ | λ (filter time), n (order) |
| SIMC (Skogestad) | $e^{(-sL)}/(1 + s\theta)$; FOTD with Tcl = θ | None (Tcl = process delay θ) |
| Pole placement (critically damped) | $\omega_n^2/(s + \omega_n)^2$ | Ts (settling time) |
| Analytical PI [16] | $p^n/(s + p)^n$ family; universal Ms, Mt, PM, GM | None (closed-form from plant) |
| **PID-SRCF** | **Any user-specified G_desired(s)** | **Target response shape** |

Under this view, PID-SRCF subsumes the listed methods rather than competing with them: setting G_desired(s) to the target row of any classical method recovers the corresponding gains up to the numerical solver tolerance, as verified by Sections 4.1–4.3 and Table 4. The contribution of PID-SRCF is therefore the explicit decoupling of the design specification (the desired closed-loop shape) from the analytical or empirical formulas that historically conflated specification and solution. The unified formulation also exposes the implicit specification of each classical method, making their relative merits comparable in a single language.

## 5. Conclusion

The PID-SRCF method presented in this study provides an optimization approach for PID controllers tuning by minimizing the $L_2$-norm error while explicitly shaping the closed-loop transient response. The method accepts any target desired step response but typically chosen for industrial relevance either as a first-order plus time delay (FOTD) system or as a second-order system with specified settling time and overshoot, offering flexibility across a wide range of control applications.

The algorithm is implemented as an open-source MATLAB script utilizing constrained nonlinear optimization via *fmincon*, ensuring stability and non-negative controller gains. This approach eliminates the need for heuristic approximations, making it suitable for both simple and complex process dynamics. Case studies on first-, second-, and third-order systems confirm the method's robustness and consistency in meeting design requirements. Comparative evaluations demonstrate that PID-SRCF can replace conventional tuning methods, like Ziegler Nichols, Lambda tuning, Pole Placement, Dominant Pole method and including MATLAB's proprietary *Control System Tuner* application.